\documentclass{caosp}

\usepackage{graphicx}
\usepackage{natbib}
\def\url#1{#1}
\articleNo{308}
\pubyear{2021}
\volume{51}
\volnumber{2}
\firstpage{1}
\received{December 04, 2020}
\accepted{February 11, 2021}

\begin{document}

\hauthor{S.V.\,Antipin {\it et al.}}

\title{New SU UMa-type star ZTF18abdlzhd in the Zwicky Transient Facility data}

\author{
       S.V.\,Antipin\inst{1}
     \and
       A.M.\,Zubareva\inst{2,1}
     \and
       A.A.\,Belinski\inst{1}
     \and
       M.A.\,Burlak\inst{1}
     \and
       N.P.\,Ikonnikova\inst{1}
     \and
       K.L.\,Malanchev\inst{3,1}
     \and
       M.V.\,Kornilov\inst{1,4}
     \and
       E.O.\,Mishin\inst{1}
       }

\institute{
           Sternberg Astronomical Institute, M.V. Lomonosov Moscow State University, Universitetskij pr., 13, Moscow, 119234, Russia, \email{serge\_ant@inbox.ru}
         \and 
           Institute of Astronomy of the Russian Academy of Sciences, Pyatnitskaya~str.,~48, Moscow, 119017, Russia
         \and 
          Department of Astronomy, University of Illinois at Urbana-Champaign, 1002 West Green Street, Urbana, IL 61801, USA
         \and
          National Research University Higher School of Economics, Staraya Basmannaya str., 21/4, Moscow, 105066, Russia
          }

\date{December 04, 2020}

\maketitle

\begin{abstract}
We carried out a search for unknown dwarf novae in a public data release  of the Zwicky Transient Facility survey and suspected that the object ZTF18abdlzhd is a SU UMa-type star. Performed multicolor CCD observations permit us to follow its fading from an outburst in August and an entire superoutburst in October 2020. The duration of the superoutburst is 13 days. We detected superhumps with period P = 0$\fd$06918(3) that are characteristic of UGSU type stars.
\keywords{stars -- photometry -- dwarf novae -- ZTF18abdlzhd}
\end{abstract}

\section{Introduction}
\label{intr}
Cataclysmic variables provide opportunities to observe various accretion-related phenomena in a setting of close binary systems. 
A white dwarf primary is receiving matter via the inner Lagrangian point (L1) from a low-mass secondary star that fills up its Roche lobe. Dwarf novae are a subclass of cataclysmic variables where the stream of matter forms an accretion disk around the white dwarf and the disk alters between hot (high accretion rate) and cold (low accretion rate) states resulting in recurrent dramatic changes in brightness. 

SU UMa-type dwarf novae are known to increase their brightness by several magnitudes for days, after which they get back to the quiescent state for a while. Two kinds of outbursts are found in these systems, distinguished by the outburst amplitude and duration. Short low-amplitude outbursts are called normal, and long ones having greater amplitudes and a ``plateau'' phase, are known as superoutbursts.  In the course of a superoutburst, periodic brightness variations called ``superhumps'' emerge. They are characterized by amplitudes up to 0.3\,mag and have periods longer than the orbital period of a system by several percent. See more detailed information on Dwarf Novae in \citet{1995CAS....28.....W}, and specifically on UGSU-type stars in \citet{2009PASJ...61S.395K}.

We revealed the object in Zwicky Transient Facility Data Release~3 \citep{2019PASP..131a8002B} during the dedicated search of dwarf novae.
We downloaded ZTF DR3 light curves from the IRSA IPAC server\footnote{\url{https://irsa.ipac.caltech.edu/data/ZTF/lc\_dr3/}}.
Then we selected all $zr$ (ZTF Sloan $r$ band, see Fig. 2 from \citet{2019PASP..131a8002B}) light curves with the following restrictions: the peak magnitude is brighter than 19.5\,mag, the amplitude is at least 1\,mag, the duration is between 10 and 30 days, and the number of observed nights is at least 10. Only good weather-condition observations were considered. This search yielded us 15 out of 3~billion ZTF DR3 objects.

A visual analysis of light curves was performed with the SNAD ZTF web-viewer\footnote{\url{https://ztf.snad.space}}
(Malanchev et al., in prep.\footnote{\url{https://arxiv.org/abs/2012.01419}}) and the object ZTF18abdlzhd was found as the only reliable UGSU candidate.

\begin{figure}
\centerline{\includegraphics[width=0.9\textwidth,clip=]{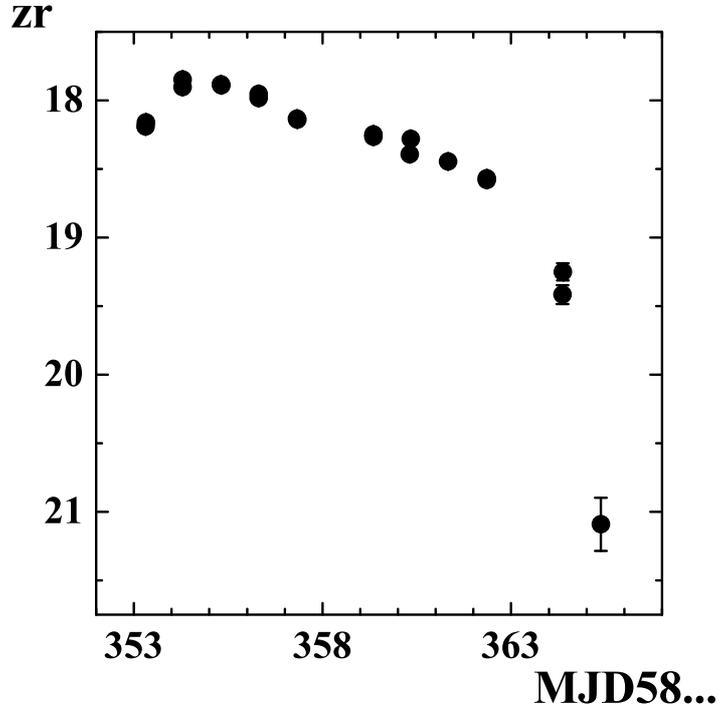}}
\caption{ZTF18abdlzhd outburst light curve from ZTF survey data. zr is for ZTF $r$ band whose transmission curve is depicted in Fig.2 in \citet{2019PASP..131a8002B}.}
\label{f1}
\end{figure}

The outburst light curve based on ZTF data is shown in Fig.\,\ref{f1}.

\section{Observations}
\label{obs}

CCD photometry of ZTF18abdlzhd ($\alpha = 23^h 15^m 30\fs841, \delta = +57\degr 08\arcmin 46\farcs98,$ J2000) carried out in August and October 2020 permit us to follow the end of the fading stage of one outburst and a full superoutburst of this dwarf nova. We monitored the field of the variable with the new automated 60-cm telescope of the Caucasus Mountain Observatory of SAI MSU. The ASA RC600 60-cm reflector was installed at the Caucasus observatory supported by M.V.\,Lo\-mo\-no\-sov Moscow State University Program of Development. It is equipped with an Andor iKon-L (DZ936N-BV) 2048~$\times$~2048 CCD-camera and a set of filters \citep[see][]{2020ARep...64..310B}.  Sloan $g$, $r$, $i$ \citep{1996AJ....111.1748F} and Johnson-Cusins $B$, $V$, $R_c$, $I_c$ bands were used to explore the behaviour of ZTF18abdlzhd. Observations in $g$, $r$, $i$ were obtained on August 19--23, 2020 (JD2459081--85, 4 nights), and CCD frames in $B$, $V$, $R_c$, $I_c$ were taken on October 10--26, 2020 (JD2459133--149, 15 nights). The observational log is given in Table~\ref{tab:obs}.
The exposure times varied from 120 seconds during the outburst to 600 seconds in minimum brightness of the object.

\begin{table}
\begin{center}
\caption{Observational log}
\label{tab:obs}
\begin{tabular}{lccll}
\hline\hline\noalign{\smallskip}
JD & Date & Number of frames (Band) \\
\hline\noalign{\smallskip}

2459081 & August 19, 2020 & 3 ($g$), 23 ($r$), 2 ($i$) \\
2459082 & August 20, 2020 & 1 ($g$), 1 ($r$), 1 ($i$) \\
2459084 & August 22, 2020 & 1 ($r$) \\
2459085 & August 23, 2020 & 1 ($r$) \\
2459133 & October 10, 2020 & 5 ($B$), 5 ($V$), 5 ($R_c$), 5 ($I_c$) \\
2459134 & October 11, 2020 & 2 ($B$), 2 ($V$), 2 ($R_c$), 2 ($I_c$) \\
2459135 & October 12, 2020 & 2 ($B$), 2 ($V$), 2 ($R_c$), 2 ($I_c$) \\
2459136 & October 13, 2020 & 2 ($B$), 2 ($V$), 2 ($R_c$), 2 ($I_c$) \\
2459137 & October 14, 2020 & 1 ($B$), 1 ($V$), 73 ($R_c$), 1 ($I_c$) \\
2459138 & October 15, 2020 & 45 ($R_c$) \\
2459139 & October 16, 2020 & 43 ($R_c$) \\
2459140 & October 17, 2020 & 2 ($B$), 2 ($V$), 2 ($R_c$), 2 ($I_c$) \\
2459141 & October 18, 2020 & 2 ($B$), 2 ($V$), 2 ($R_c$), 2 ($I_c$) \\
2459144 & October 21, 2020 & 1 ($B$), 1 ($V$), 1 ($R_c$), 1 ($I_c$) \\
2459145 & October 22, 2020 & 1 ($B$) \\
2459146 & October 23, 2020 & 6 ($R_c$) \\
2459147 & October 24, 2020 & 1 ($R_c$) \\
2459148 & October 25, 2020 & 1 ($R_c$) \\
2459149 & October 26, 2020 & 1 ($R_c$) \\
\noalign{\smallskip}\hline\hline
\end{tabular}
\end{center}
\end{table}

To perform aperture photometry and magnitude calibration, we used 
VaST\footnote{\url{https://scan.sai.msu.ru/vast}} software \citep{2018A&C....22...28S}. We derived magnitudes of an ensemble of comparison stars within the field of view from the APASS ($B$, $V$, $R_c$, $I_c$) and the PanSTARRS1 ($g$, $r$, $i$) survey \citep{2016arXiv161205560C}.

\section{Results}
\label{res}

\begin{figure}[thp]
\centerline{\includegraphics[width=0.41\textwidth,clip=]{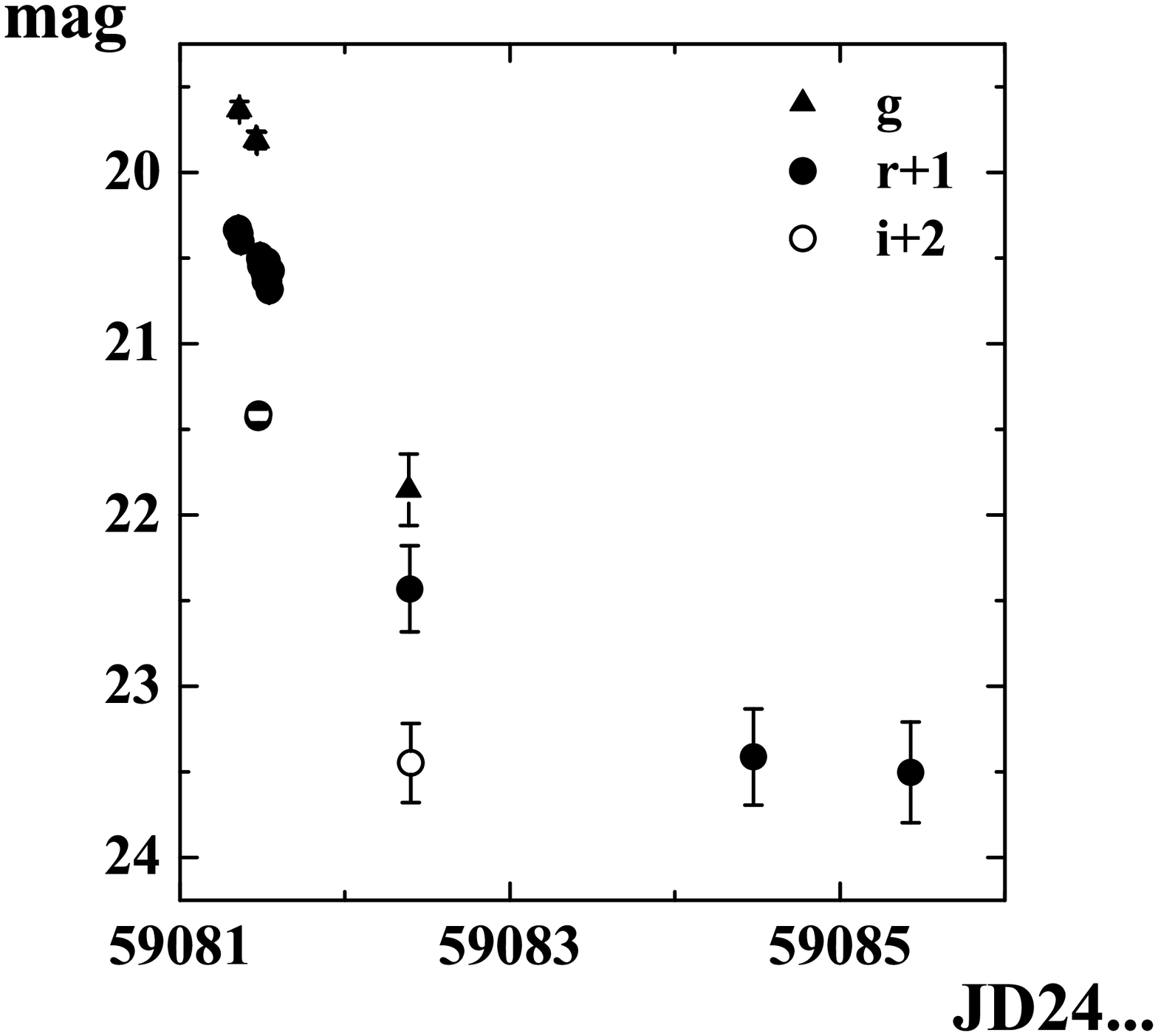} 
           \includegraphics[width=0.59\textwidth,clip=]{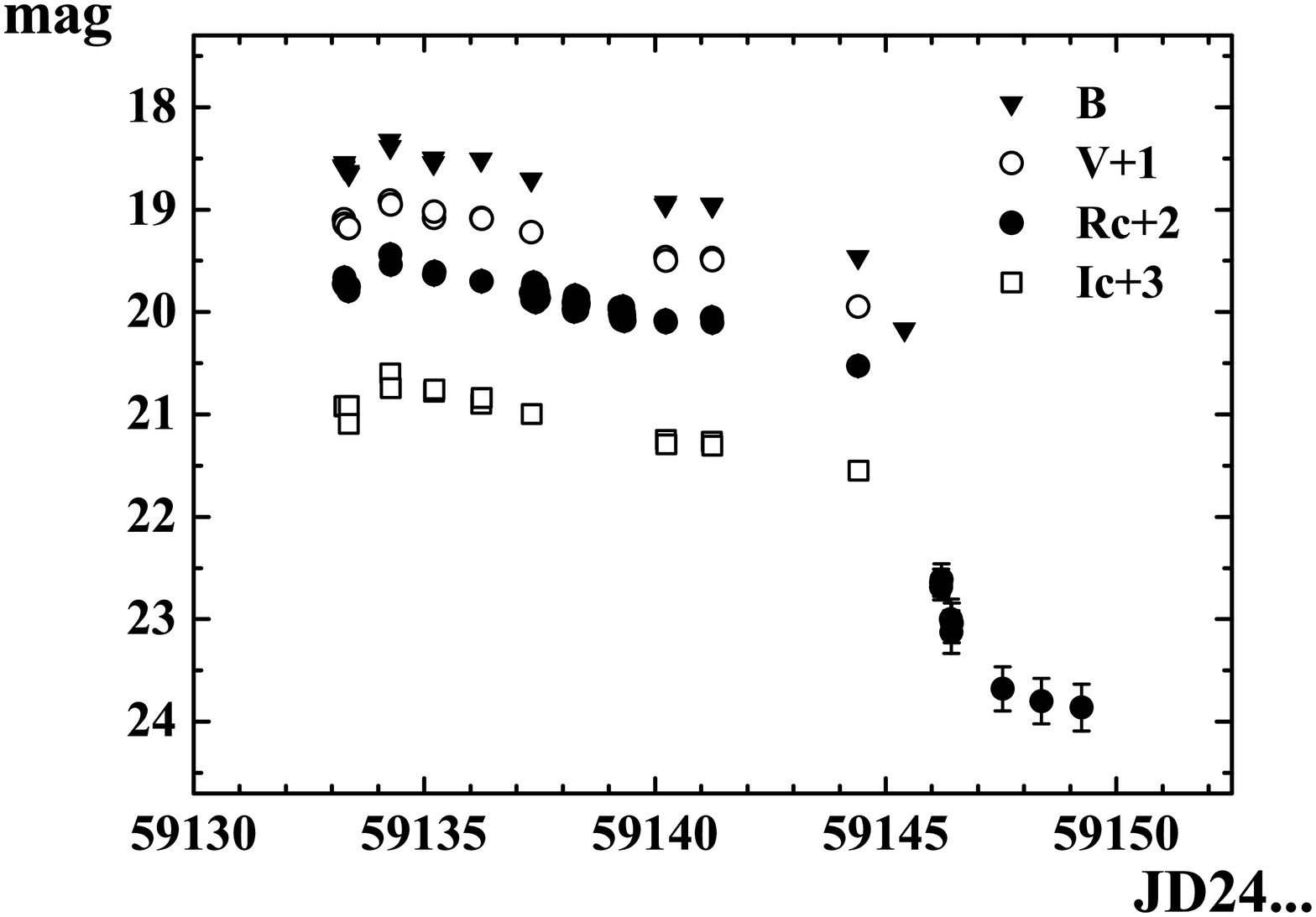}
           }
\caption{ZTF18abdlzhd. The fading stage of the August outburst (left panel) and the superoutburst of October 2020 (right panel).}
\label{f2}
\end{figure}

Our data cover two outbursts of ZTF18abdlzhd. Unfortunately, we have no opportunity to classify the type of the August outburst (normal or superoutburst, see the left panel of Fig.\,\ref{f2}) because the observations reveal only the final stage of the event.

On October 10, 2020, we found the variable in its bright state again. The brightness of the variable star increased to the next night of observations, so we can say that we detected the dwarf nova at the beginning of an outburst. The total duration of this outburst is 13 days which is very similar to the event found in ZTF data (cf. Fig.\,\ref{f1} and the right panel of Fig.\,\ref{f2}).

ZTF18abdlzhd reached the maximum brightness on October, 11, $R_c = 17.49$ mag. Immediately after the superoutburst, the star faded to $R_c = 21.8$\,mag. In the next few days, the dwarf nova passed beyond the detection limit (about 22.0\,mag in $R_c$-band at 600 s exposure). Thus, the variability amplitude is greater than 4.5\,mag. We found ZTF18abdlzhd in PanSTARRS DR1 (object ID 176573488784746366) at $r$ = 22.096 $\pm$ 0.196\,mag \citep{2016arXiv161205560C}).

Color indices remain constant during the plateau phase of the outburst and are equal to $B-V$ = 0.46 $\pm$ 0.02, $V-R_c$ = 0.41 $\pm$ 0.02 and $V-I_c$ = 0.23 $\pm$ 0.03. 

\begin{figure}
\centerline{\includegraphics[width=1\textwidth,clip=]{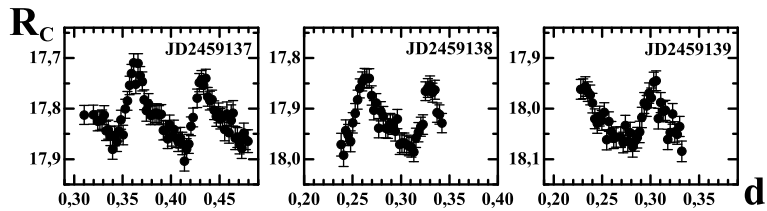}}
\caption{Individual light curves of ZTF18abdlzhd for three nights in maximum of the superoutburst (October 14, 15 and 16, 2020).}
\label{f3}
\end{figure}

\begin{figure}[thp]
\centerline{\includegraphics[width=0.5\textwidth,clip=]{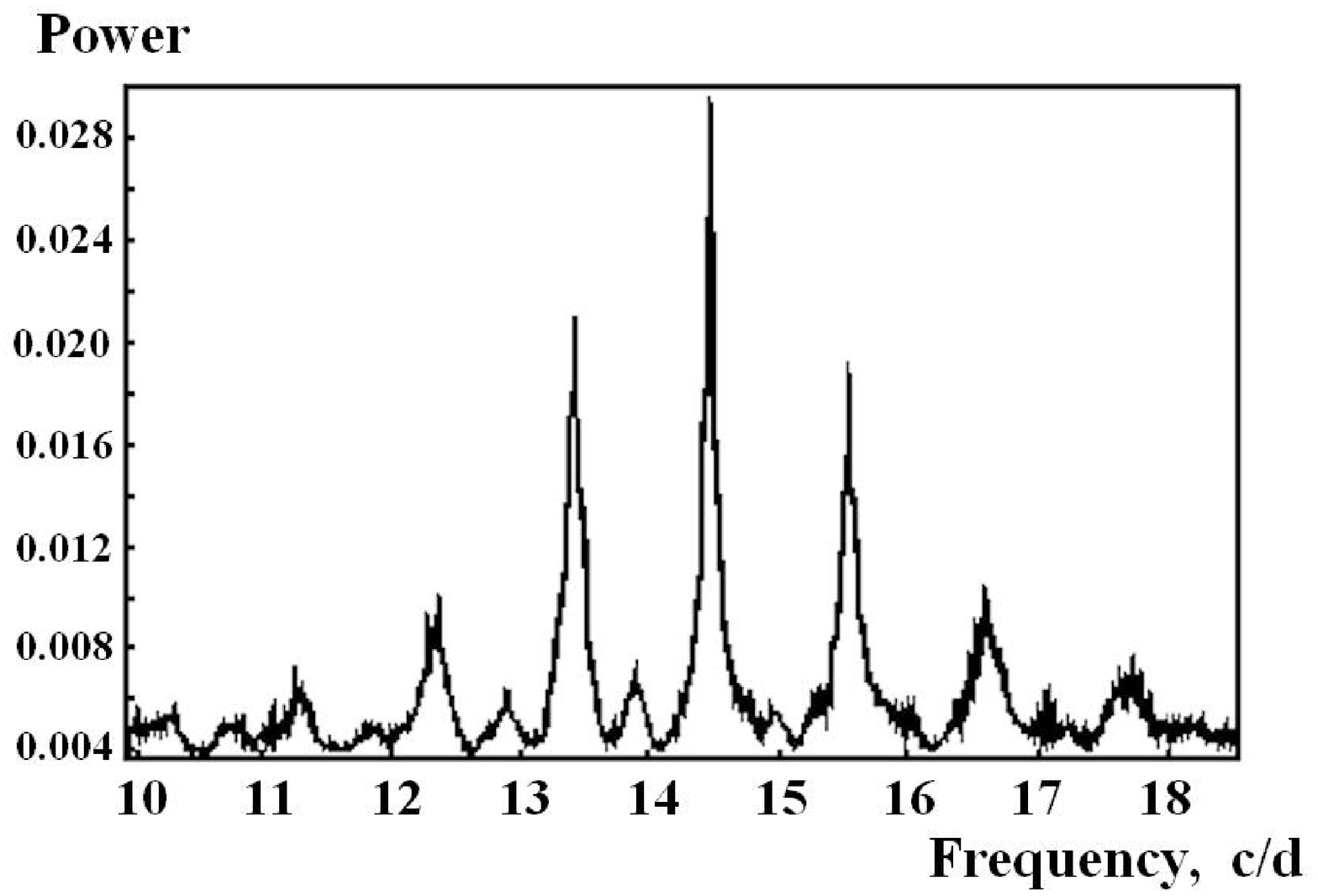} 
           \includegraphics[width=0.5\textwidth,clip=]{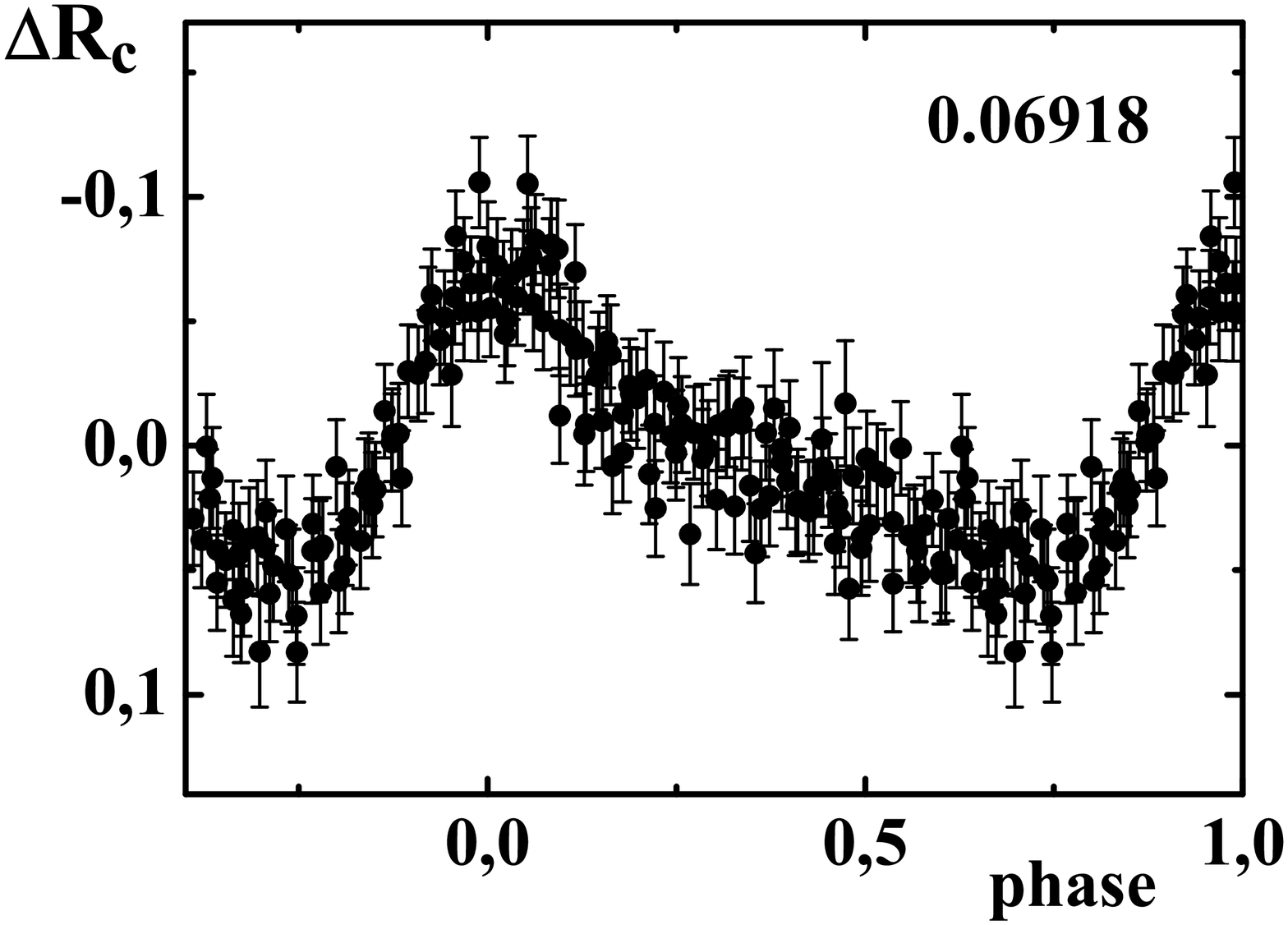}
           }
\caption{Left panel: the periodogram for three nights of observations shown in Fig.\,\ref{f3}. Maximum peak corresponds to the frequency 14.455(6) c/d and period 0.06918(3) days. Right panel: the corresponding phased light curve of superhumps.}
\label{f4}
\end{figure}

During the maximum of the October 2020 outburst we found a periodic brightness variability -- superhumps -- which are a distinctive feature of a superoutburst (see Fig.\,\ref{f3}). The most of our observations in the plateau phase of this superoutburst were obtained in $R_c$ band -- 161 frames for three nights covered with photometry densely (see Table~\ref{tab:obs}). For this set we removed a linear trend from the brightness measurements before running the period search for which WinEfk software\footnote{\url{http://www.vgoranskij.net/software/}} developed by Dr. V.P.~Goranskij was applied. The corresponding periodogram and the phased light curve of the superhumps are given in Fig.\,\ref{f4}. The ephemeris is as follows:

$$ JD_{max} = 2459137.4314 + 0.06918(3)\times E.   $$

\section{Conclusion}
\label{concl}

Based on the ZTF survey data, we suspected that object ZTF18abdlzhd is a UGSU-type star. Our CCD observations allow us to confirm this assumption. The duration of the October 2020 superoutburst was 13 days. We found a superhump period P = 0$\fd$06918(3).

\acknowledgements
The authors are grateful for partial support from M.V.\,Lo\-mo\-no\-sov Moscow State University Program of Development.
This research has been supported by the Interdisciplinary Scientific and Educational School of Moscow University ``Fundamental and Applied Space Research''.
K.L.M. and M.V.K. are supported by RFBR grant 20-02-00779 for the preparing ZTF data and the revealing of the object.
The study was made possible through the use of the AAVSO Photometric All-Sky Survey (APASS), funded by the Robert Martin Ayers Sciences Fund and NSF AST-1412587.

\bibliography{ztf}

\end{document}